\documentstyle[11pt,newpasp,twoside,epsf]{article}
\def\edcomment#1{\iffalse\marginpar{\raggedright\sl#1\/}\else\relax\fi}
\reversemarginpar

\begin{document}
\title{Maps of the Cosmos: The Cosmic Microwave Background}
 \author{Lyman A. Page}
\affil{Princeton University, Dept. of Physics,
Jadwin Hall, Washington Rd}

\begin{abstract}
Since the IAU XXIV meeting in 2000, the CMB anisotropy has matured from being
one of a number of cosmological probes to forming the bedrock foundation
for what is now the standard model of cosmology.
The large advances over the past three years have come from
making better and better maps of the cosmos. 
We review the state of measurements of the anisotropy and 
outline some of what we have learned since 2000.
The recent advancements may be placed roughly into three categories:
1) What we learn from the CMB with minimal input from
other cosmic measurements such as the Hubble constant;
2) What we learn from the CMB in combination with other probes
of large scale structure; and 3) What we learn by using
the CMB as a back light. Future directions are also discussed.
It is clear: we have much more to learn from the CMB anisotropy.
\end{abstract}
 
\section{Introduction}

It has long been appreciated that the CMB anisotropy could be a powerful
probe of cosmology. The foundations
of the anisotropy calculations we do today were set out
over thirty years ago by Sachs \& Wolfe (1967), 
Rees (1968), Silk (1968), Peebles \& Yu (1970), and
Sunyaev \& Zeldovich (1970). Plots of the acoustic peaks
were shown in Doroskevich, Zeldovich, \& Sunyaev (1978) 
and Bond \& Efstathiou (1984) gave the results of detailed numerical
calculations.  

On the measurement side, the tension between expectations and continuously
improving upper limits (e.g., Weiss 1980, Wilkinson 1985, 
Partridge 1995) was finally alleviated by the discovery of the 
anisotropy by {\sl COBE} (Smoot et al. 1992).  At that time,
the measured Sachs-Wolfe plateau ($l<20$) was a factor of two {\it higher}
than expectations based on the standard cold dark matter model in
which $\Omega_m\approx 1.$\footnote{We use the convention that
$\Omega_m=\Omega_{cdm}+\Omega_b+\Omega_\nu$ is the cosmic density in 
all matter components where $cdm$ is cold dark matter, $b$ is for baryons,
and $\nu$ is for neutrinos;  $\Omega_r$ is the cosmic radiation
density (now minuscule); $\Omega_\Lambda$ is the corresponding 
density for a cosmological 
constant; and $\Omega_k$ is the corresponding curvature parameter.
The Friedmann equation tells us: 
$1\equiv\Omega_\Lambda+\Omega_k+\Omega_m=\Omega_{tot}+\Omega_k$. The
physical densities are given by, for example, $\omega_b=\Omega_bh^2$.}
There were many measurements of 
the anisotropy at the {\sl COBE} scales and finer between 1992 and 2000
(e.g., see Page 1997 for a table) that culminated in 
observations of the first acoustic peak (Dodelson \& Knox 2000, 
Hu 2000, Pierpaoli, Scott \& White 2000, 
Knox \& Page 2000). 
 
Amidst theories that did not survive observational tests and false clues,
a standard cosmological model emerged. Even over a decade ago, 
the evidence from a  majority of independent tests indicated 
$\Omega_m\approx0.3$ (e.g., Ostriker 1993). It was realized 
by many that a flat model ($\Omega_k=0$) with a significant 
cosmic constituent with negative pressure, such as a cosmological constant,
was a good fit to the data. Then, in 1998 measurements of type 1a 
supernovae (Riess et al. 1998, Perlmutter et al. 1999) 
directly gave strong indications that the universe was
accelerating as would be expected from a cosmological constant. 
The state of the observations in 1999 is summarized in 
Figure~1. If the Einstein/Friedmann equations describe our universe,
the data were telling us that the universe is spatially flat 
with matter density $\Omega_m\approx 0.3$ and 
$\Omega_{\Lambda}\approx 0.7$. Independent analyses came to similar
conclusions (e.g., Lineweaver 1998, Tegmark \& Zaldarriaga 2000). 
The story since the last IAU meeting is that the concordance
model does in fact describe virtually all cosmological
observations astonishing well. There is now a well agreed upon 
standard cosmological model (Spergel et al. 2003, 
Freedman \& Turner 2003).   

\begin{figure}
\plotfiddle{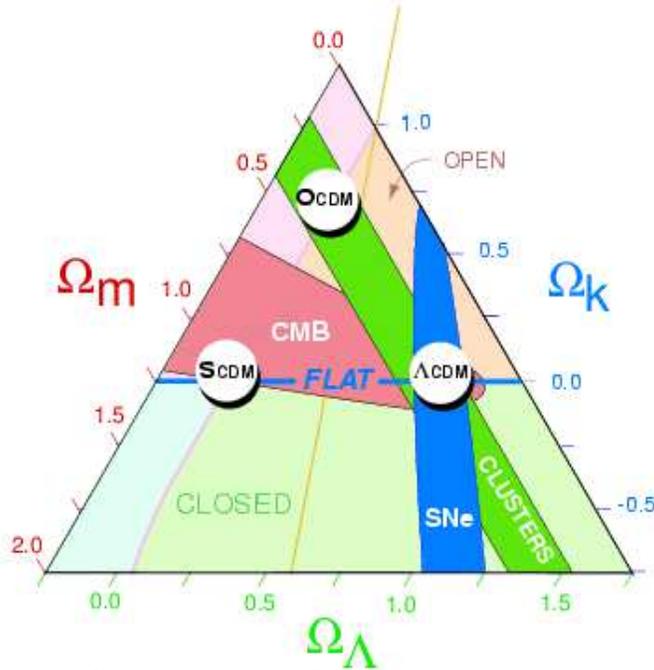}{3.2truein}{0}{70}{70}{-120}{-20}
\caption{The Cosmic Triangle from Bahcall et al. (1999).
This shows the concordance 
model for cosmological observations at the end of the last millenium. Three
different classes of observations, supernovae, clusters, and CMB anisotropy,
are consistent if we live in a spatially flat universe with a cosmological 
constant.}
\label{fig:triangle}
\end{figure}

In the rest of this article we briefly review the CMB observations 
in \S2 and summarize what we learn from (almost) just 
the CMB in \S3. We then, in \S4, outline 
the things we learn by combining the CMB with other maps of the 
cosmos, in particular the 2dF Galaxy Redshift Survey (Colles et al. 2001).
In \S5 we indicate the sorts of things we hope to learn by using the CMB as
a back light for lower redshift phenomena. We conclude in  \S6. 

\section{CMB Observations \label{sec:cmbobs}}

To be sure there have been advances in all cosmological observations
over the past three years, but the most dramatic improvements have
come from observations of the CMB anisotropy. First came the BOOMERanG results
which gave us high resolution and high signal-to-noise maps of the 
anisotropy (deBernardis et al. 2000). The data analysis improved
considerably over the three years (e.g., Netterfield et al. 2002)
culminating in the Ruhl et al. (2003) analysis. Concurrent with BOOMERanG
was the MAXIMA experiment (Hanany 2000, Lee 2001). MAXIMA gave similar results 
though over a smaller fraction of the sky. For many, the greatest 
advance from these experiments was not so much the new measurement of 
curvature, but rather the ability to probe $\omega_b$ and $\omega_m$
with better than 30\% precision using the CMB. 
In more general terms, they gave new and strong evidence that
the concordance model in Figure 1 was correct. 

In the year before the {\sl WMAP} release (Bennett et al. 2003), 
the Archeops team 
published (Benoit 2002)  results from
a map that covered roughly 30\% of the sky. The goal of the experiment,
in addition to serving as a test bed for the Planck HFI instrument,
was to bridge the angular range from the $7^\circ$ {\sl COBE} resolution
to $1^\circ$ resolution. The ACBAR experiment (Kuo et al. 2002), 
done from the South Pole,
was aimed at pushing to angular scales 
beyond what {\sl WMAP} could reach. Its resolution is $0.08^\circ$ 
as opposed to {\sl WMAP}'s $0.21^{\circ}$. BOOMERanG, MAXIMA, 
Archeops, and ACBAR achieved
their greatest sensitivity at 150 GHz using radiometers based on
the Berkeley/Caltech/JPL spiderweb bolometers (Bock et al. 1996)
with passbands defined by filters developed by Peter Ade and colleagues
at Cardiff.

Great strides were made in CMB interferometry during the past three
years. The three primary instruments were DASI (Halverson et al. 2002), 
VSA (Grainge 2003), and CBI (Mason 2003, Pearson 2003). 
All were based on broadband 30 GHz HEMT amplifiers designed 
by Marian Pospieszalski at the National Radio Astronomy Observatory
(Pospieszalski 1992). Results from DASI first complemented 
and extended the framework 
that was becoming evident. After adding the polarization capability, 
DASI discovered the intrinsic polarization in the CMB at the predicted
level (Kovac et al. 2002). This was an important piece of evidence 
that decoupling occurred as predicted. The VSA interferometer gave 
similar results to DASI though over a wider range in $l$. 
The CBI interferometer
clearly observed the suppression of the anisotropy at $l>1000$
due to Silk damping and the finite thickness of the decoupling
surface. CBI also showed hints of observing the formation of 
non-linear structure at $l>2000$, though more investigation is
needed as emphasized by the CBI team.

Though the advances since IAU XXIV by ground and balloon based 
CMB experiments were tremendous, the results from {\sl WMAP}
are in a different category. Not only did {\sl WMAP} have 
the unprecedented stability achievable only from deep space, but
it mapped the entire sky. The systematic error limits achieved on 
multiple different aspects of the experiment and analysis were roughly
an order of magnitude (sometimes two orders) improvement over what
had been achieved previously. The data are so clean that 99\% of the time
ordered data goes into the final map. There is a low level of filtering
and a 1\% transmission imbalance is corrected, but other than this
no other sytematic error corrections
or selection criteria are applied. Finally, all the 
data from the experiment are publicly available so they may be checked.
For a description of the {\sl WMAP} mission see the article by 
Chuck Bennett in these proceedings. 

The new CMB observations have narrowed the CMB 
swath in Figure~1 by roughly an order of magnitude.
More importantly, they have told us that adiabatic scale invariant
fluctuations seeded the formation of cosmic structure and that the contents
of the universe are baryons, some form(s) of dark matter, and some
form(s) of dark energy. A snapshot of all the CMB anisotropy data
as of July 2003 has been compiled by Bond, Contaldi, \& Pogosyan (2003) 
and a version is shown as a Grand Unified Spectrum (GUS) in Figure~2. 

\begin{figure}
\plotone{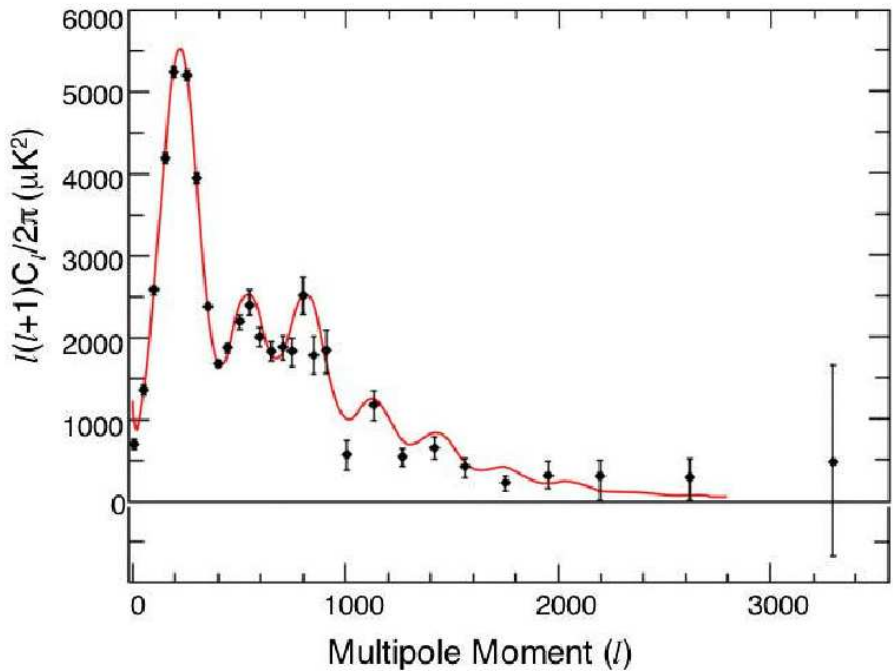}
\caption{The Grand Unified Spectrum based on 
Bond, Contaldi, \& Pogosyan (2003).
The y axis shows the fluctuation power per logarithmic interval in $l$.
The x-axis may be converted to angular scale by $\theta_{deg}=l/200$.
This spectrum is derived from the combination 
of 28 anisotropy experiments as of July 2003. 
The first and second peaks from the acoustic
oscillations are clearly evident, the third peak is almost 
resolved, and the damping tail at $l>1000$ is evident. 
The line is the best fit model.} 
\label{fig:gus}
\end{figure}

This IAU is a particularly good time to take stock of where we are.
Another chapter in the study of the CMB has been finished with
the release of the first year {\sl WMAP} data. Sadly, Dave Wilkinson,
a pioneer of the CMB field for 35 years and a founder of both the 
{\sl WMAP} and {\sl COBE} satellite missions, died in the end of 2002 after 
battling cancer for 17 years. Fortunately Dave saw the {\sl WMAP} maps 
in their full glory. The {\sl MAP} satellite was renamed in his honor.
Figure~3 is from the {\sl WMAP} launch
and shows three of CMB science's pioneers.
 
\begin{figure}
\plotfiddle{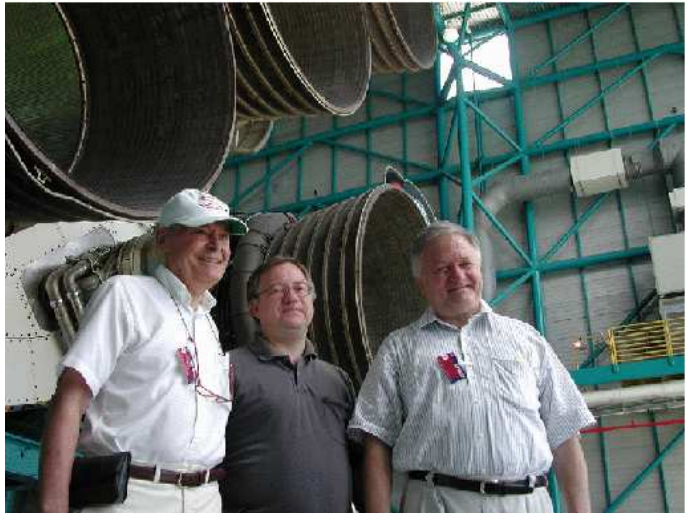}{3.2truein}{0}{50}{50}{-130}{0}
\caption{David Wilkinson (left), Dick Bond (center) and Rashid Sunyaev
(right) at the {\sl WMAP} launch, June 2001. They are standing in front 
of a Saturn V rocket ({\sl WMAP} used a Delta II). 
Prof. Rashid Sunyaev was the recipient of the 2003 Gruber Prize in 
Cosmology which was presented at the IAU symposium.}
\label{fig:dtw_db_rs}
\end{figure}

\section{What the CMB data alone tell us.}
\label{sec:cmbpar}

As a good first approximation, one should think of a map of the CMB
anisotropy as a picture of the universe at a redshift of $z_{dec}=1089$,
when the CMB decoupled from the primordial plasma. 
Thus, the CMB tells us about the universe when it was less
than $t_{dec}=379$~kyrs old and a much simpler place. In this epoch, 
the early universe acts as though it is spatially flat, independent
of the values of the dark energy and dark matter today. 

The variation in 
temperature from spot to spot across the sky arises from the primordial 
plasma responding to spatial variations in the gravitational potential.
In turn, the gravitational landscape is the manifestation of quantum
fluctuations in some primordial field. In the inflationary model,
one imagines these fluctuations stretched by at least $10^{28}$ 
so that they are super-horizon size, and then expanded with the 
expansion of the universe.

Observing the CMB is like looking at a distant 
surface\footnote{This is the ``surface'' at which
the CMB decoupled from the primordial electrons and baryons. It is
sometimes called the last scattering surface, but since 
$\approx$15\% of the CMB photons were really last 
scattered when the universe was reionized near $z=20$, we prefer decoupling.} 
at the edge of the observable universe. As the universe expands,
the pattern in the anisotropy will shift as new regions of
the gravitational landscape are sampled. For example, one may imagine
that the quadrupole ($l=2$) may rotate $90^{\circ}$ in one Hubble
time (30~mas/century), with higher multipoles changing faster.
In a similar vein, the light from the clusters of galaxies that formed in 
the potential wells that gave rise to cold regions on the decoupling
surface has not has enough time to reach us.  

The processes of the formation of stars, galaxies, and clusters of
galaxies takes place between us and the decoupling surface. As a first 
approximation, photons from the decoupling surface come to us 
unimpeded. The lower redshift properties do, though, affect the light from the
decoupling surface but in characteristic and definable ways as
discussed below.

A full analysis of the CMB involves accurately comparing 
the measured power spectrum, Figure~2, to models.
The simplest model that describes the CMB data is flat
and lambda-dominated. The results for this parametrization
derived from {\sl WMAP} alone (Spergel et al. 2003) and  
the independent GUS analysis are shown in Table~1. 

\begin{table}
\begin{center}
\caption{Cosmic Parameters from CMB measurements}
\begin{tabular}{lcccc} 
\hline
Description              & Parameter      & {\sl WMAP}   & GUS & w/2dF\\
\hline
Baryon density  & $\Omega_bh^2$   & $0.024\pm0.001$ 
                         & $0.023\pm0.002$ & $0.023\pm0.001$ \\
Matter density  & $\Omega_mh^2$   & $0.14\pm0.02$  
                         & $0.14\pm0.01$   & $0.134\pm0.006$\\
Hubble parameter  & h                      & $0.72\pm0.05$  
                         &  $0.71\pm0.05$  & $0.73\pm0.03$\\
Amplitude  & A  & $0.9\pm0.1$  &   $0.85\pm0.06$ & $0.8\pm0.1$ \\
Spectral index  & $n_s$  & $0.99\pm0.04$   &  $0.967\pm0.029$ 
                         & $0.97\pm0.03$   \\
Optical depth  & $\tau$  & $0.166^{+0.076}_{-0.071}$  & $\cdots$   
                         & $0.148\pm0.072$ \\
\end{tabular}
\end{center}
\label{tab:cosparm}
\end{table}
We can get at the essence of what the CMB is telling us from the 
following. Let us focus on the decoupling surface. There is 
a natural length scale
in the early universe that is smaller than the horizon size. 
It corresponds to the distance over which a density perturbation 
(sound wave) in the primordial plasma can propagate in the age of 
the universe at the time of decoupling ($t_{dec}=379~$kyr). It is called 
the acoustic horizon. Once we know the contents of the universe from 
the overall characteristics of the power spectrum, we can compute
the size of the acoustic horizon. It is 
roughly $r_s\approx c_s t_{dec}z_{dec}$
where $c_s$ is the sound speed in the plasma. In the full expression
(Hu \& Sugiyama 1995), $r_s$ depends on only the physical densities
of matter and radiation and not on the Hubble parameter, h.
We may think of $r_s$ as a standard yard stick embedded in the 
decoupling surface. From a map of the anisotropy, we measure the angular
size, $\theta_A$, of the feature corresponding to $r_s$. 
From {\sl WMAP}, $\theta_A=0.598^{\circ}\pm0.002$.
By definition then,

\begin{equation}
\theta_A\equiv{ r_s(z_{dec})\over d_A(z_{dec}) }
\end{equation}
where $d_A$ is the angular size distance to the decoupling surface.
In $d_A$ we can trade off the geometry, 
$\Omega_k=1-\Omega_r-\Omega_m-\Omega_{\Lambda}$,
with $h$. Thus to determine the geometry without recourse to appealing 
to the simplest model, we must make a prior assumption on $h$.
The dependence is not strong. If one assumes $h>0.5$ then 
one finds $0.98<\Omega_{tot}<1.08$ (95\% cl), 
where again we have used the {\sl WMAP} data
for illustration. The progress in our knowledge of $\Omega_{tot}$
as determined by all available data roughly between the past two IAU
symposia (starting with Figure~1, Bond et al. 2003) is:

\begin{table}
\begin{center}
\caption{Total Cosmic Density, $\Omega_{tot}$ ($1\sigma$ errors)}
\begin{tabular}{lc} \hline
January 2000   & $\Omega_{tot}=1.06^{+0.16}_{-0.10}$\\
January 2002   & $\Omega_{tot}=1.035^{+0.043}_{-0.046}$\\
January 2003   & $\Omega_{tot}=1.034^{+0.040}_{-0.042}$\\
March 2003 (+{\sl WMAP})  &  $\Omega_{tot}=1.015^{+0.063}_{-0.015}$ \\   
\end{tabular}
\end{center}
\label{tab:omegas}
\end{table}

One way to see what the CMB alone can tell us is to plot the data 
in the $\Omega_m-\Omega_\Lambda$ plane for a pure cosmological
constant, or equation of state $w=-1$. This is shown in 
Figure~4 for the {\sl WMAP} data.
All simple open, flat, and closed
cosmological models satisfying the Friedmann equation can 
be plotted here. One picks a point in the space, a single
source of the fluctuations (e.g., adiabatic fluctuations in the metric from
an inflationary epoch), $w=-1$, and marginalizes over the other
parameters ($n_s$, $\omega_b$, $\tau$, $A$ ) with uniform priors. 
The possibilities are labeled by the Hubble parameter that goes with them.

There are a number of things the plot pulls together. First,
there is a large degeneracy in the CMB data along the line that
runs above the line for flat universe. This
is called the ``geometric degeneracy'' and is essentially the observation 
noted above that one must pick $h$ to determine $d_A$ to complete
the equation $\theta_A=r_s/d_A$. The degeneracy line clearly misses a
model in which the universe is flat with $\Omega_m=1$ 
($\Omega_\Lambda=0$), the 
Einstein-deSitter case. If one stretches the data slightly, it is possible
to have a model with $\Omega_m\approx 1.3$ ($\Omega_\Lambda=0$) but the
price one pays is a Hubble parameter near 0.3. This value is in conflict
with a host of other non-CMB observations. In addition, when one considers the 
Integrated Sachs-Wolfe (ISW) induced cross-correlation between cosmic 
structure, as measured by radio sources, and the CMB anisotropy, 
this solution is disfavored at the $3\sigma$ level (Nolta et al. 2003).
Thus, in this minimal picture, there are no models with $\Omega_\Lambda=0$
that fit the data.
 
Once one moves off the x axis, the intersection of the 
flat universe line, $\Omega_\Lambda+\Omega_m=1$, and the geometric degeneracy
is the next least baroque point, at least by today's standards of 
baroqueness.
It is very satisfying that h for the intersection is very close
to the value obtained from the Hubble Key Project 
($h=0.72\pm0.03(stat)\pm0.07(sys)$, Freedman et al. 2001).
Additionally, the values agree with probes of the large scale structure
and the supernovae data. From the plot, it is easy to see why
such a weak prior on h (or $\Omega_m$) picks out a flat universe.
A number have noted that all determinations of $\Omega_{tot}$ are
greater than unity. The plot shows that with the priors we have chosen,
there are more solutions with $\Omega_{tot}>1$. This may bias the solution 
somewhat. 

\begin{figure}
\plotfiddle{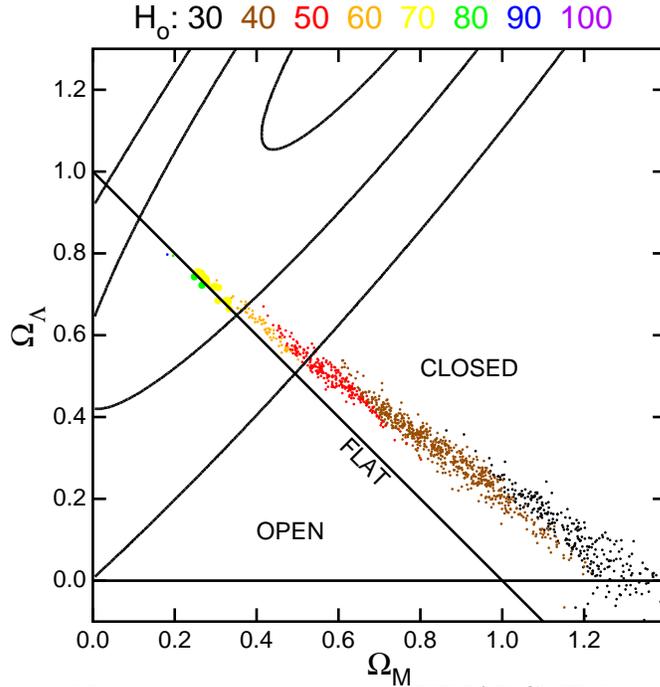}{3.2truein}{0}{50}{50}{-150}{-90}
\caption{Models consistent with the {\sl WMAP} CMB data in the 
$\Omega_\Lambda-\Omega_m$ plane. The flat models correspond to the line
with  $\Omega_\Lambda+\Omega_m=1$. This plot assumes that the dark energy has
$w=-1$. The code at the top gives the values of the Hubble constant
as one moves along the geometric degeneracy. It is striking that
the value picked out by the CMB for a flat universe, $h=0.71$, is in such
agreement with the value from the HST key project. The observations 
behind these two probes are completely different and correspond
to times separated by a good fraction the age of the universe.
The $1\sigma$, $2\sigma$, and $3\sigma$ contours for the
supernovae are plotted as well (Tonry et al. 2003). 
Constraints from large scale structure would correspond to 
roughly a vertical swath centered on
$\Omega_m=0.3$. This plot is courtesy of Ned Wright.}
\label{fig:olamom}
\end{figure}

\section{The CMB in combination with other cosmic probes}
\label{sec:cmblss}

We learn much more about cosmology when we add to the CMB anisotropy
lower redshift observations. The primary CMB anisotropy comes from a 
surface behind the galaxies and clusters of galaxies which  
are roughly between us and $z\approx2$ (or between now, $t_U=13.7$~Gyr,
and when the universe was 3.3~Gyr). 
In regards to cosmic parameters, lower redshift measurements 
sample the universe in a much different state of it evolution
and therefore with different parameter degeneracies. In regards
to understanding structure formation, the CMB gives us the initial 
conditions whereas the lower redshift
measurements of the large scale structure (LSS) give us the 
current condition.

There are a number of ways in which lower $z$ measurements complement
the CMB: (a) through measuring the current expansion rate
with the Hubble constant; (b) through measuring the current baryon 
density with quasar
absorption systems; (c) through measuring the current mass density
with galaxy velocities or the mass-to-light ratio; 
(d) through measuring the ages of the 
oldest objects; and (e) through measuring the matter power spectrum with 
gravitational lensing (e.g., Contaldi, Hoekstra, \& Lewis 2003) 
or galaxy surveys such as 2dFGRS and 
SDSS (e.g., Gunn et al. 1998).\footnote{The SDSS/{\sl WMAP} analysis came 
out after the IAU.}

The complementarity of the Hubble constant has been discussed
above; and the other probes, of course, have a rich history. Since 
the last IAU, huge strides were made in determining the matter power 
spectrum as we discuss below. The supernovae results are not 
included in the list because 
it still seems best to treat the CMB+LSS as an independent probe of
negative pressure.

The power spectrum from galaxies, $P_g(k)$, and the CMB power spectrum are
intimately related. However, technical issues arise when comparing the two
because one is not certain how fluctuation in the number density
of galaxies trace the fluctuations in matter. In other words, 
the galaxy population might be {\it biased} with respect to the 
matter density which the CMB probes.
The bias is quantified as $P_g(k)=b^2P(k)$ where $P(k)$
is the matter power spectrum and $b$ is 
the bias factor. For example, it is observed that redder (e.g., IRAS)
galaxies cluster together less strongly than 
do optically selected ones and are thus less 
strongly biased (Fisher et al. 1994). Similarly, luminous galaxies 
are more biased than less luminous ones (Norberg et al. 2001). 

The amplitude of the matter power spectrum is set by $\sigma_8$,
the rms fluctuations in the matter mass density in a 
comoving sphere of diameter 8~Mpc. In order to determine $\sigma_8$,
one needs to know the cosmic matter density, $\Omega_{m}$, 
which is only determined to 30\% accuracy (as opposed to the physical
matter density which is determined to 
15\% accuracy). In the CMB, $\sigma_8^2$ simply scales the overall
amplitude of the angular power spectrum, whereas $\omega_m$
is sensitive to the shape of the spectrum. 

Figure 5 shows a comparison of the CMB power spectrum and $P(k)$. 
The big leap in galaxy surveys in the past few years 
is that the 2dFGRS survey  was able to 
measure the matter power precisely and over a large range in $k$,
particularly values of $k$ directly probed by the CMB. The two data sets 
are combined by comparing their shapes and amplitudes as discussed in 
Verde et al. (2003).

The primary observables that 2dF adds are the extended baseline
over which the fluctuations are measured and an independent measure
of the dark matter power spectrum. (A value of the bias of 
$b=1.06\pm0.11$ (Verde et al. 2002) was used in the {\sl WMAP} 
analysis.) These break a number of 
parameter degeneracies inherent in just $l<1000$ CMB measurements.
For example, it is clear that with a longer baseline in $k$ the
spectral index, or overall tilt of the spectrum, can be better 
determined. It is then easier to determine the optical depth, $\tau$,
and the matter density $\omega_m$.
In Table~1, one can see the improvement in what one can say 
when the 2dF data are added to the CMB data (WMAP+ACBAR+CBI,
Spergel et al. 2003).

\begin{figure}
\plotfiddle{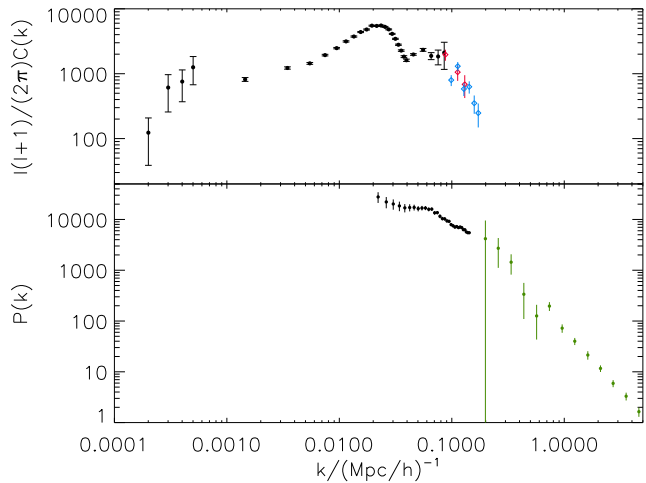}{3.2truein}{0}{50}{50}{-150}{0}
\caption{Top: the CMB angular power spectrum as a 
function of comoving wavenumber $k$ ($k\approx l/14400$). The 
points to the left
are from {\sl WMAP}, the data for $l>875$ are from CBI and ACBAR.
Bottom: the LSS data from 2dFGRS ($0.01<k<0.15$) and the Lyman$\alpha$
survey (Croft et al. 2002). The power spectra have been rescaled to $z=0$.
To compare the LSS data to the CMB on must take into account
redshift-space distortions, non-linearities, 
bias, window functions,  etc. This
figure from Verde et al. (2003)}
\label{fig:cmblss}
\end{figure}

The high-z/low-z combination leads to other science as well.
With the emergence of these precise probes, one can now constrain 
the neutrino mass at the levels being probed by particle physics 
experiments. The phenomena is as follows. When they are 
relativistic, neutrinos free-stream out of the potential wells.
Because neutrinos are relativistic at early times 
and cool with the expansion of the universe, clustering
by neutrinos is suppressed on small scales with respect to large scales.
As one increases the cosmic mass density of neutrinos, $\Omega_\nu$,
while holding $\Omega_m=\Omega_{cdm}+\Omega_{\nu}+\Omega_b$ fixed,
the net matter fluctuations, $P(k)$, are also suppressed on small
scales (high $k$). At the same time, the length scale of the 
suppression moves to smaller values (higher $k$). By comparing the
$P(k)$ from 2dF with the {\sl WMAP} CMB fluctuations, one finds 
$\Omega_\nu h^2<0.007$ (95\% cl) or $\Sigma {m_\nu}<0.7~$eV (95\% cl) 
for three degenerate neutrino species.
In a sense, we have started addressing questions of
classic particle physics with cosmological probes.

There are also direct correlations expected between the CMB anisotropy 
and galaxy surveys in addition to the relation between power spectra. 
Crittenden \& Turok (1996) pointed
out that in a $\Omega_\Lambda$ dominated universe, there should be 
measurable correlations between the CMB anisotropy and the 
matter as traced by x-ray or radio sources at large angular scales.
The mechanism is that the gravitational potential wells change, due to the 
$\Lambda$-induced acceleration, while a CMB photon traverses it. This
in turn affects the energy of the photons (ISW effect). These same
potential wells are traced by galaxy populations around $z\sim2$.
The correlation has been seen albeit at only about the $3\sigma$ level
(Boughn \& Crittenden 2003, Nolta et al. 2003, Fosalba, Gaztananga 
\& Castander 2003, Scranton et al. 2003).

The standard model has many predictions and consistency checks
that are currently being tested. For the CMB, a number of the 
correlations that should exist are given 
in Peiris \& Spergel (2000). Of course, the pay dirt is in the
inconsistencies!

\section{The CMB as a back light}
\label{sec:cmbbl}

The physical processes that gave rise to
cosmic structure leave characteristic and identifiable 
signatures on the CMB that can be seen by using the CMB as a
back light. The key aspects of the CMB as an illumination
source are that we know the redshift at which the fluctuations
were imprinted and we know the frequency spectrum to high accuracy.
Probably the best known examples of using the CMB as a back light
are the Sunyaev-Zeldovich (SZ) effects (1972). These were discussed 
in a companion session by John Carlstrom and so I'll not discuss 
them here. Instead, I'll focus on the formation of the first stars 
and on gravitational lensing.

The process of formation of the first stars is not well understood
and not well observationally constrained. However, it is known that 
the intergalactic hydrogen in the universe was predominantly neutral 
after decoupling and is predominantly ionized now. It was the formation 
of the first stars that reionized the universe. The free electrons
from the reionization leave an imprint on the CMB. It can be shown
that scattering by an electron in a quadrupolar radiation background
polarizes the CMB. Because this happens at low redshifts, $z\approx20$,
the CMB appears polarized at large angular scales. This effect was seen 
in the first year {\sl WMAP} data through the polarization-temperature
correlation (Kogut et al. 2003).

With the large angular scale anisotropy, one measures the amount
of polarized emission and directly infers an optical depth to 
polarization, $\tau$. The most likely value from WMAP is 
$\tau=0.17\pm0.04$. In other words, roughly 15\% of the CMB
photons were rescattered by the formation of the first 
stars (Zaldarriaga 1997). The redshift of $z\approx20$ is obtained
by integrating back over a completely ionized universe until
$\tau=0.17$ is reached. This corresponds to an age of
200~million years after the bang. There is still much more work to 
be done to understand the ionization history of the universe.

The reionization suppresses the CMB fluctuations at medium scales
but gives rise to a new fluctuations at smaller angular scales
through the Ostriker-Vishniac effect (OV, Ostriker \& Vishniac 1986). 
The physics is similar to
that of the kinetic SZ effect though is applied to
density perturbations instead of clusters per se.
In other words, the CMB photons are scattered by ionized gas
with some peculiar velocity.
The effect has the frequency spectrum of the CMB and must be 
separated from the primary anisotropy through spatial filtering
and higher order statistics. Its measurement is one of the goals
of the next generation of experiments. Not only is it of 
intrinsic interest, but it also will be a new handle on breaking the 
$n_s-\tau$ cosmic parameter degeneracy.

The CMB is lensed, like distant galaxies, by the intervening mass 
distribution. A picture of this is shown in 
Figure~6. The effect of the lensing is large but it will 
challenging to separate the intrinsic CMB from what we measure
(the lensed CMB). The pursuit is worthwhile because from the lensing
one can extract $P(k)$ without bias 
(Seljak \& Zaldarriaga 1999, Okamoto \& Hu 2003).
To a first approximation, 
lensing redistributes the phase of the anisotropy and so the 
power spectra of the lensed and unlensed sky are the same.
However, there is also a net redistribution of power and so 
lensing enhances the angular power spectrum at high $l$.

There are two avenues to detecting lensing. One is through 
higher order statistics (e.g., the four point function, 
Bernardeau 1997). Because lensing distorts the intrinsic 
hot and cold spots, a lensed sky has more complicated statistics
than the two-point function that describes the intrinsic anisotropy.
One can get a sense for this from
Figure~6: the difference map has
elongated features. Detection through this method requires 
a high fidelity map. The other avenue is through the
CMB polarization. Lensing distorts the E-mode CMB polarization 
from the decoupling surface, producing B-modes (Zaldarriaga \& Seljak 1998). 
Indeed, it is the largest
B-mode signal at $l>200$. In a sense, one uses the polarization to
filter out the lensing signal from the intrinsic CMB and OV effects. 

A goal for cosmologists over the next few years is to use these 
probes together to study the growth of cosmic structure as a 
function of redshift. For example, reionization tells us
what's happening at $z\approx20$, the OV effects
and diffuse thermal SZ probe the early stages of structure formation,
the kinetic and thermal SZ effects in clusters, and the lensing
of the CMB probe the later stages of structure formation.
By combining these probes, one estimates that  
the equation of state, $w$, can be measured
to 10\% accuracy and the neutrino mass
to 0.1~eV. The new thing is that these determinations will be tied to the 
CMB and will not rely on galaxy surveys. In addition, there is a 
rich set of correlations and cross 
checks between various measurements, both within the CMB and between 
the CMB and optical lensing, that should permit us to build confidence
in any conclusions we may draw. 

\begin{figure}
\plotfiddle{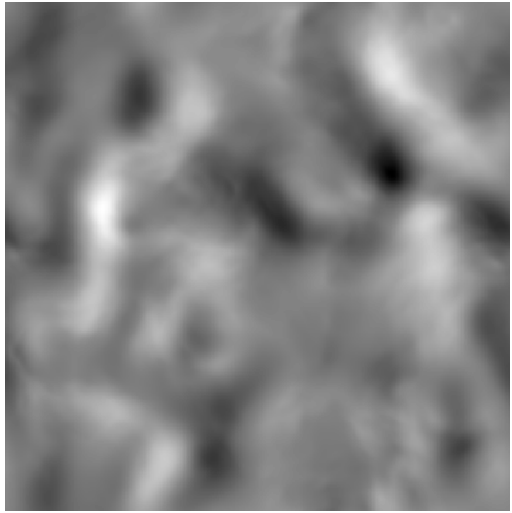}{3.0truein}{0}{50}{50}{-100}{0}
\caption{The difference between a lensed and unlensed CMB field.
The image is $1^{\circ}$ on a side. The rms amplitude is a few $\mu$K
with several peaks reaching 20-30~$\mu$K. Note the coherence in the lensed
features. Figure courtesy of Uros Seljak.}
\end{figure}

\section{Concluding Remarks}
\label{sec:cmbcon}

Over the past few years, and especially with {\sl WMAP}, the CMB data has
become the foundation for the standard cosmological model.
Any model that purports to explain the birth and evolution of the
universe must be able to predict the results in Figure~2. 
This is a very stringent requirement. The model
elements implicit in the figure---superhorizon fluctuations  
with cosmic structure seeded by a scale invariant spectrum with
Gaussian fluctuations in the metric are at
the core of our conception of the universe. They are also at the 
heart of inflation. Indeed, we have started to directly constrain
models of inflation (Peirie et al. 2003).
This is not to say that our currently favored 
model is correct. There are elements of the observations, for example the 
apparent suppression of fluctuations on the largest angular scales,
that may call for something beyond the standard model.
However, it is truly astounding that we have a model that naturally 
explains almost all cosmological observations. The model is eminently
testable and precise enough to be experimentally challenged. The 
model is also young enough to admit new discoveries in such areas
as dark energy, dark matter, and the birth and growth of cosmic structure.

We have much more to learn from the CMB. To borrow from 
Winston Churchill, {\sl WMAP} marks not the end, not even the beginning of the 
end, but rather the end of the beginning of what we can learn from the CMB. 
In IAUs ahead we may hope to hear of how observations of the CMB
in combination with other cosmic probes determine the mass of 
the neutrino or the equation of state of the dark energy. Detection
of polarization B-modes (See A. Couray, these proceedings) may be 
be able to tell us the energy scale of inflation. From the ground,
new experiments such as ACT, APEX, and SPT are pushing CMB anisotropy
measurements to high $l$ and high sensitivity. New experiments
such as BICEP, CAPMAP, Polarbear, QUAD, {\sl SPORT}, are applying 
new techniques
to measure the polarization in the CMB. There is already talk of 
{\sl CMBPOL}, a post-Planck satellite dedicated to polarization measurements.
No doubt, precise measurements of the CMB will continue to shed 
light on fundamental physics, cosmology, and astrophysics for years to come. 

\acknowledgments

The synopsis above benefitted greatly from discussions
with the WMAP team, Dick Bond, Arthur Kosowsky, Uros Seljak 
and Suzanne Staggs. Figure 4 is from Ned Wright and Figure 6 is 
from Uros Seljak. We thank Dick Bond and colleagues for sharing their 
compilation and analysis before publication.

%
%
%

\end{document}